\documentstyle[epsfig,longtable,lscape,physics_old_fm]{aipproc}

\newcommand{\syst}{\mbox{(\text{syst})}}

\newcommand{\mrad}{\mbox{$\mathrm{\,mrad}$}}

\newcommand{\Bmeson}{\mbox{$\mathrm{B}$}}

\newcommand{\BmesonUD}{\mbox{$\mathrm{B}_{u,d}$}}

\newcommand{\BorBstar}{\mbox{$\mathrm{B}^{(*)}$}}

\newcommand{\BorBstarp}{\mbox{$\mathrm{B}^{(*)+}$}}
\newcommand{\Bdstar}{\mbox{$\mathrm{B}^{**}$}}

\newcommand{\BdstarUD}{\mbox{$\mathrm{B}^{**}_{u,d}$}}

\newcommand{\Btwostar}{\mbox{$\mathrm{B}_2^{*}$}}
\newcommand{\Bone}{\mbox{$\mathrm{B_1}$}}
\newcommand{\Bonestar}{\mbox{$\mathrm{B}_1^{*}$}}
\newcommand{\Bzerostar}{\mbox{$\mathrm{B}_0^{*}$}}
\newcommand{\Bsubc}{\mbox{$\mathrm{B}_c^{+}$}}

\newcommand{\Dzero}{\mbox{$\mathrm{D^0}$}}
\newcommand{\Dstar}{\mbox{$\mathrm{D}^{*}$}}
\newcommand{\DorDstar}{\mbox{$\mathrm{D}^{(*)}$}}

\newcommand{\Dstarp}{\mbox{$\mathrm{D}^{*+}$}}
\newcommand{\Ddstar}{\mbox{$\mathrm{D}^{**}$}}
\newcommand{\Ddstarzero}{\mbox{$\mathrm{D}^{**0}$}}

\newcommand{\DdstarSone}{\mbox{$\mathrm{D}^{+}_{s1}$}}
\newcommand{\DdstarStwo}{\mbox{$\mathrm{D}^{*+}_{s2}$}}
\newcommand{\Dtwostar}{\mbox{$\mathrm{D}_2^{*}$}}
\newcommand{\Dtwostarzero}{\mbox{$\mathrm{D}_2^{*0}$}}
\newcommand{\Done}{\mbox{$\mathrm{D_1}$}}

\newcommand{\Dstarpr}{\mbox{$\mathrm{D}^{*'}$}}

\begin{document}
\title{B and D Spectroscopy at LEP}

\author{Franz Muheim$^*$}
\address{$^*$
Universit\'e de Gen\`eve   \\
D\'epartement de physique nucl\'eaire et corpusculaire\\
24, quai Ernest-Ansermet, 
1211 Gen\`eve 4, Switzerland \\
Email: Franz.Muheim@cern.ch  \\
presented at Heavy Quarks at Fixed Target HQ98 workshop, Fermilab, Oct. 1998.}

\begin{flushright}UGVA-DPNC 1998/11-180\\
Nov. 1998
\end{flushright}

\vspace*{-1cm}
\maketitle

\begin{abstract}
Results from the four LEP experiments ALEPH, DELPHI, L3, and OPAL on
the spectroscopy of B and charmed mesons are presented. The  predictions of 
Heavy Quark Effective Theory (HQET) for the masses and the widths of excited $L=1$
B mesons are supported by a new measurement from L3. A few \Bsubc\ candidate events
have masses consistent with the recent CDF observation and the predictions.
New results on \Ddstar\ production and 
$\Bmeson \to \Ddstar \ell \nu$ are also presented. The evidence for a \Dstarpr\ meson 
reported recently  by DELPHI is not supported by OPAL and CLEO.

\end{abstract}

\section*{Introduction}

Detailed understanding of the spectroscopy of orbitally excited heavy  mesons containing 
a $b$ or a $c$ quark 
provides important information regarding the underlying theory.
A flavor-spin symmetry arises from the fact that the mass of a  heavy quark $Q$ 
is large
relative to $\Lambda_{\mathrm{QCD}}$.  In this approximation, the spin
$\vec{s}_Q$ of the heavy quark $Q$ is conserved 
in the interactions, independently of the total
angular momentum $\vec{j}_q = \vec{s}_q + \vec{L}$ of the light quark $q$.
Corrections to this symmetry are a series expansion in $1/m_Q , 1/m_Q^2$, calculable in
Heavy Quark Effective Theory (HQET) \cite{HQS}.

\begin{table}[!htb]
    \caption{ $L=1$  B mesons containing a $u$ or a $d$ light quark with
corresponding spin states, relative production rates, prediction for masses and widths, and two-body decay modes.\label{tab:decays}}
\begin{center}
    \begin{tabular}{lccccccc}
      {\bf Name } & {\bf $j_q$} & {\bf $J^P$} & {\bf Production} & {\bf Mass [MeV]} 
	& {\bf Width [MeV]} &\multicolumn{2}{c}{\bf Decay Mode} \\ [1mm]
\tableline
      \Bzerostar & $1/2$ & $0^+$ & 1/12 &$M_{\Bonestar}-12$  & $\sim 150$ &$\Bmeson\pi$ & S-wave \\
      \Bonestar  & $1/2$ & $1^+$ & 3/12 &$M_{\Btwostar}\pm 100$ &$\sim 150$&$\Bstar\pi$ & S-wave \\
      \Bone      & $3/2$ & $1^+$ & 3/12 &5759 & 21 & $\Bstar\pi$                & D-wave \\
      \Btwostar  & $3/2$ & $2^+$ & 5/12 &5771 & 25 & $\Bstar\pi,\Bmeson\pi$ & D-wave \\ [1mm]
    \end{tabular}
  \end{center}
\end{table}

The $L=0$ mesons, for which $j_q = 1/2$, have two possible
spin states: a pseudo-scalar $P$  ($J^P = 0^-$) and a vector $V$ ($J^P = 1^-$).  
If the spin of the heavy quark is 
conserved independently, the relative production rate of these states
is expected to be $V/(V+P) = 0.75$. Corrections due to the decay of
higher excited states are predicted to be small. Recent measurements of this
rate for the B system \cite{BstarL3,BstarLEP} agree well with this ratio.

In the case of  $L=1$ orbitally excited B mesons 
two sets of 
doublets are expected: the \Bzerostar\ and \Bonestar\   ($j_q = 1/2$)
and the \Bone\ and \Btwostar\ ($j_q = 3/2$) mesons (see Table~\ref{tab:decays}).
Their relative production rate follows from spin state counting (2J+1 states)\cite{SpinParity}.
For the dominant two-body decays, the $j_q = 1/2$ states
can decay via an  S-wave transition  and their decay widths are expected to
be broad in comparison to those of the $j_q = 3/2$ states which must decay via
a  D-wave transition.  
Many measurements exist for  $L=1$ orbitally excited charm mesons. 
All six narrow states, a doublet (\Dtwostar\ and  \Done) for each quark content 
($c \bar u,  c \bar d$ and  $c\bar s$) are well established\cite{Ddstar}.
The wide  $L=1$ states are hard to measure and have not been clearly identified.

Several models based on HQET 
and on the charmed $L=1$ meson data, have made 
predictions for the masses and widths of orbitally excited \Bdstar\ mesons\cite{Gronau,Eichten,Falk,Isgur,Ebert} (see Table~\ref{tab:decays}).
Some of these models place the average mass of the $j_q = 3/2$ states above that of the 
$j_q = 1/2$ states, while others predict the opposite (``spin-orbit inversion'').
The mass splitting within each doublet is predicted to be 12 MeV.

\section*{\Bdstar\ Spectroscopy}

At LEP excited states of B mesons are produced. Each of the four experiments has collected about 
$4 \times 10^6$ hadronic events out of which $0.9 \times 10^6$ events contain $B \bar B$ pairs.
In all \Bdstar\ analyses, first, the $b$-quark purity
of the data sample is increased by applying  a lifetime based event tag. 
\Bmeson\ mesons are reconstructed 
inclusively. Typically the two most energetic jets of the event are considered B candidates.  
The decay products of the B meson are separated from the background due to 
fragmentation particles using secondary vertex tagging (for charged decay particles) or the
rapidity of the decay products with respect to the B-jet axis. 
An alternative method is to fully reconstruct the B-meson decay chain which improves the 
resolution B-mass resolution but suffers from low statistics.

The decay of a \Bdstar\ meson ($\Bdstar  \to \BorBstar \pi$) 
is carried out
via a strong interaction and thus the transition pion  
originates at the primary event vertex.
In addition, the predicted masses for the $L=1$ states correspond to
relatively small $Q$ values, so that the pion direction
is forward with respect to the B-meson direction. 
The track with the largest component of momentum in the direction
of the B jet is selected.  

\begin{figure}[ht]
  \begin{center}
	\mbox{\epsfig{figure=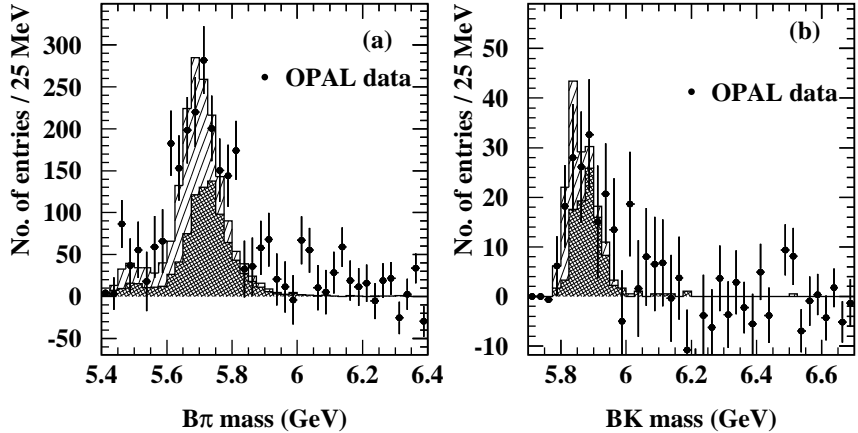,width=0.9\linewidth}}
  \end{center}
  \caption{The invariant mass distribution for (a) $\BorBstarp \pim$  and  (b) $\BorBstarp \km$ 
	combinations after subtraction of the background, respectively.
	The solid histograms shows the contribution from the \Btwostar\ and the hatched histogram 
	shows the contribution from the \Bone\ state.
            \label{fig:BdstarOPAL}}
\end{figure}

A first measurement of $L=1$ orbitally excited B mesons has been presented by OPAL\cite{BdstarOPAL}.
Using secondary vertex charge tagging to inclusively reconstruct B mesons,
the invariant mass distributions  of  $\BorBstarp \pim$ and  $\BorBstarp \km$ combinations 
show enhancements consistent with the decay of  \Bdstar\ resonances as shown in 
Fig.~\ref{fig:BdstarOPAL}.
An excess of 1738 $\pm$ 121 $\BorBstarp \pim$ and  149 $\pm$ 30 $\BorBstarp \km$ candidates 
is observed in the mass ranges 5.60 - 5.85 \GeV\ and 5.80 - 6.00 \GeV, respectively. 
The background is estimated with the wrong (like)-sign combinations.
Fitting the excess with a single Breit-Wigner function yields an average  
mass $M(\BdstarUD)$ and a production rate 
 $f_{\Bdstar} = {\cal B}( b \to \BdstarUD ) / {\cal B}( b \to \BmesonUD)$. 
Throughout this paper,
isospin symmetry is always employed to account for \Bdstar\ decays via neutral pions.
DELPHI and ALEPH have made similar measurements using rapidity to inclusively reconstruct B mesons\cite{BdstarDELPHI,BdstarALEPHi}. 
The results are summarized in Table~\ref{tab:Bdstari}.

\begin{table}[thb]
    \caption{Inclusive \Bdstar\ measurements.\label{tab:Bdstari}}
  \begin{center}
\begin{tabular}{lcc}
	& {\bf $M(\BdstarUD)$ [\MeV]} & {\bf $f_{\Bdstar}$} \\ [1mm]
\tableline
OPAL  	& $5681 \pm 11$ 		& $0.270 \pm 0.056$ \\
DELPHI 	& $5734 \pm 5 \pm 17$ 	& $0.32 \pm 0.018 \pm 0.06$ \\
ALEPH   & $5703 \pm 4 \pm 10$ 	& $0.279  \pm 0.016 \pm 0.059 \; ^{+0.039}_{-0.056}$ \\[1mm]
\end{tabular}
  \end{center}
\end{table}

\begin{figure}[hbt]
\vspace*{-0.5cm}
\begin{minipage}[h]{.49\linewidth}
  \begin{center}
  	\mbox{\epsfig{figure=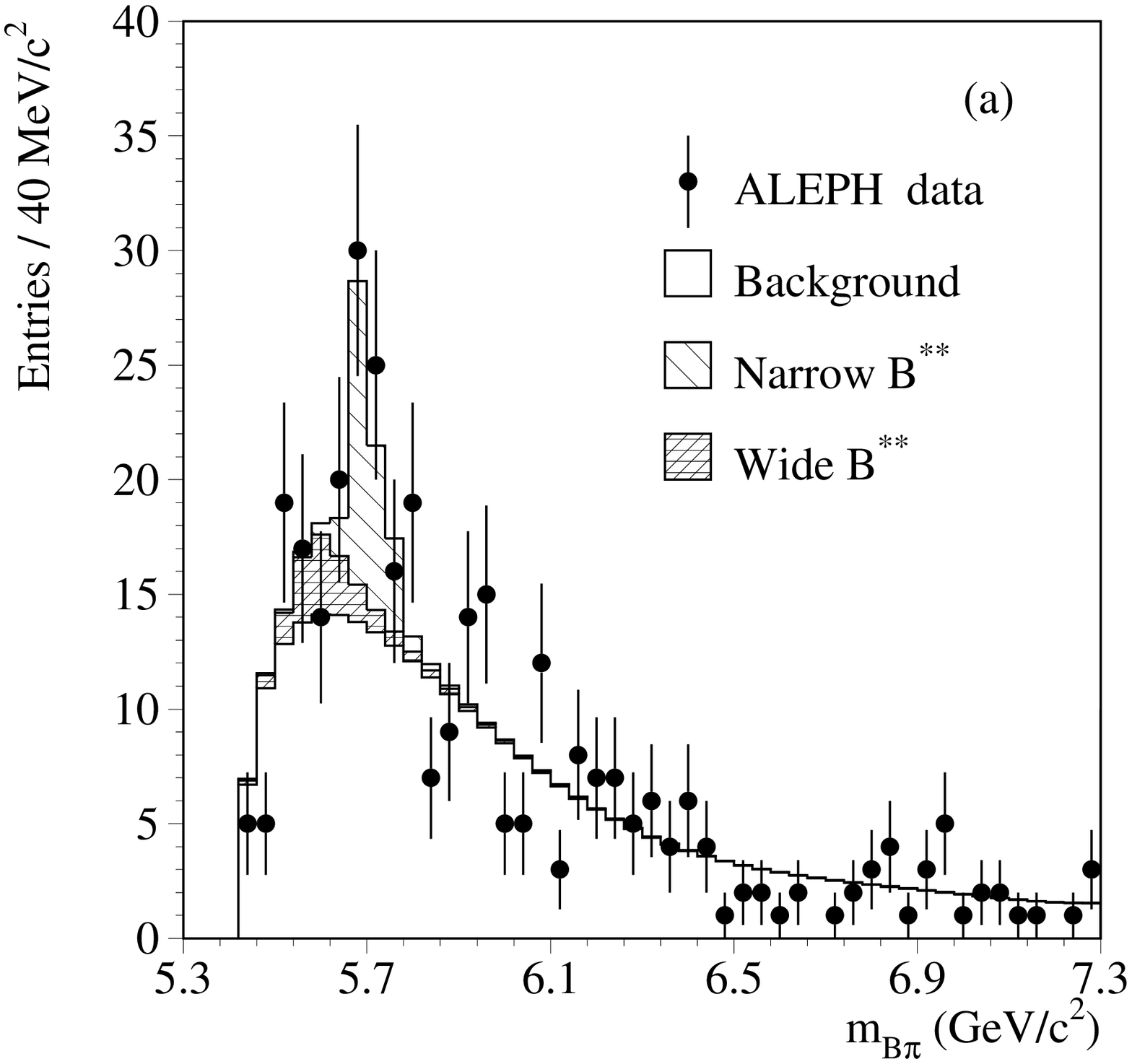,width=0.99\linewidth}}
  \end{center}
\end{minipage}
\begin{minipage}[h]{.49\linewidth}
  \begin{center}
  	\mbox{\epsfig{figure=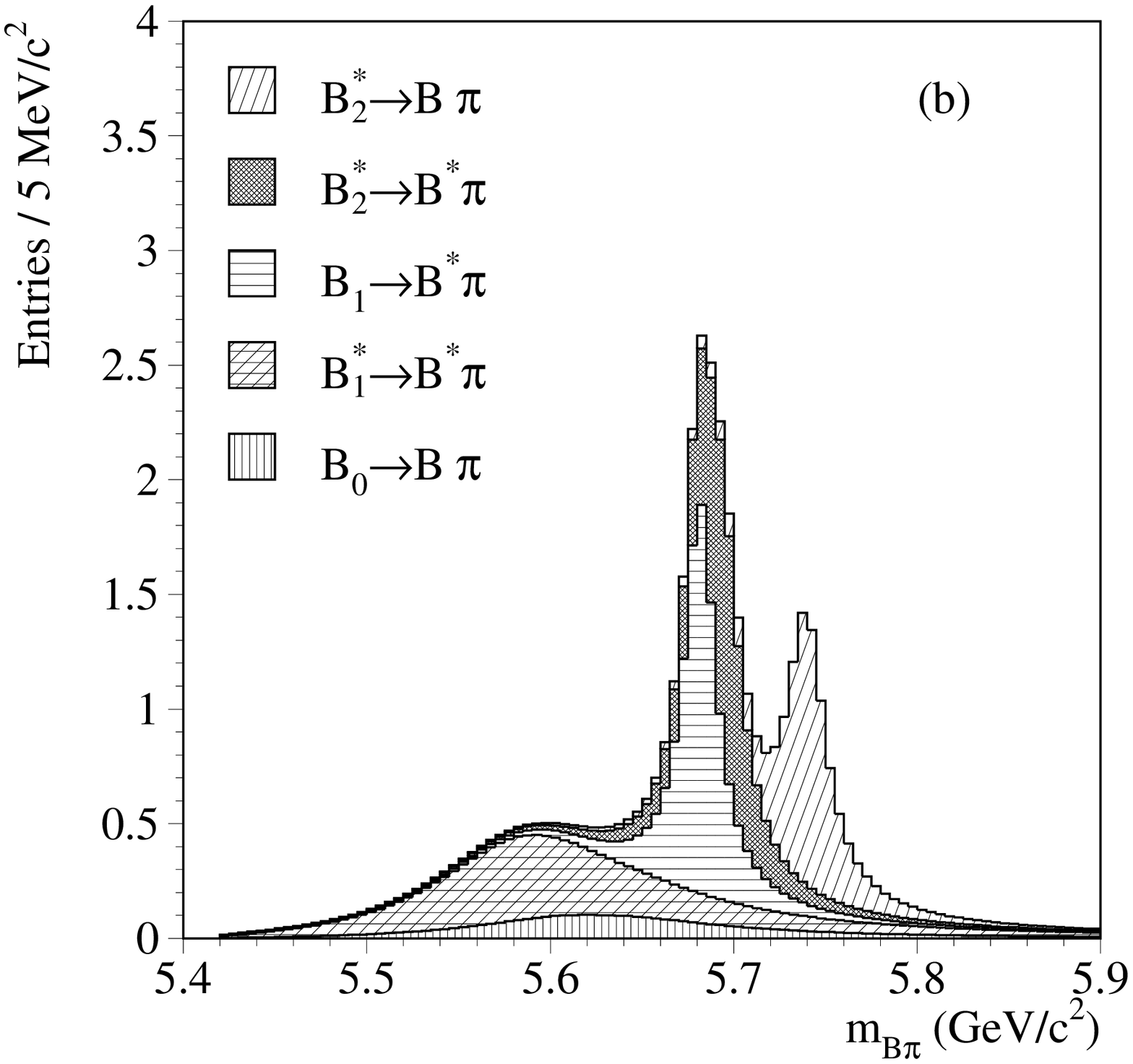,width=0.99\linewidth}}
  \end{center}
\end{minipage}
\caption{(a) B$\pi$ mass spectrum from data. The fit (histogram) includes the expected background 
plus contributions from the narrow and wide \Bdstar\ states. 
(b) An expanded view of the signal region.
  \label{fig:BdstarALEPH}}
\end{figure}

A new measurement using an exclusive method is presented by ALEPH\cite{BdstarALEPH}. 
Using many decay modes ($\Bmeson \to \DorDstar X$, where $X \in [\pi, \rho, a_1]$ and
 $\Bmeson \to J/\psi (\psip) K^{(*)}$)
238 charged and 166 neutral B meson candidates have been fully reconstructed.
The sample has a B meson purity of 85 \%.
Each  B candidate is then combined with 
a charged pion from the primary vertex.
An excess of $45 \pm 13$ events is seen in the right-sign sample compared to the wrong-sign sample.
Fig.~\ref{fig:BdstarALEPH} shows a fit to the right-sign mass spectrum
where the signal shape consists of five Breit-Wigner peaks. The relative masses, the widths, and 
the relative production rates of the individual \Bdstar\ mesons have been 
fixed to the predictions from HQET. 
The mass of the \Btwostar\ meson and the overall production rate are measured to be:
\begin{eqnarray*}
M_{\Btwostar} &=& 5739 \; ^{+8}_{-11} \; ^{+6}_{-4} \MeV \\
f_{\Bdstar} & = & 0.31 \pm 0.09 \; ^{+0.06}_{-0.05} \quad .
\end{eqnarray*}
%

A new analysis using an inclusive method is presented here by the L3 
experiment\cite{BdstarL3}. 
Several techniques are used to both improve on the 
resolution of the $\Bmeson\pi$ mass spectrum and to
unfold this resolution from the signal components.
 As a result, L3 is able to
extract measurements for the masses and widths of both the D-wave $\Btwostar$
decays and the S-wave $\Bonestar$ decays.

B meson candidates are reconstructed inclusively from all charged and neutral particles with rapidity $y > 1.6$ relative to the original jet axis. 
The measurement of the
direction of the B meson is determined
by an error-weighted average of the direction of the measured secondary decay vertex
and of the direction of the B candidate.
The angular resolution obtained is $\sigma_1 = 12~\mrad$ for  $\phi$, and  
$\sigma_1 = 18~\mrad$ for $\theta$, respectively.
The energy of the B meson candidate,  $E_{\Bmeson}$, is estimated by taking advantage of
the known center-of-mass energy at LEP, $E_{\mathrm{cm}}$, to be
\begin{equation}
  \label{eq:Benergy}
    E_{\Bmeson} = \frac{E^2_{\mathrm{cm}}-M^2_{\Bmeson}+M^2_{\mathrm{recoil}}}
                       {2E_{\mathrm{cm}}} \quad,
\end{equation}
where $M_{\mathrm{recoil}}$ is the mass of all particles in the event recoiling against the 
\Bmeson\ candidate.
The difference between
reconstructed and generated values for the B-meson energy can be described by
an asymmetric Gaussian with widths of $1.9~\GeV$ and
$2.8~\GeV$.

\begin{figure}[hbt]
  \begin{center}
    \mbox{\epsfig{file=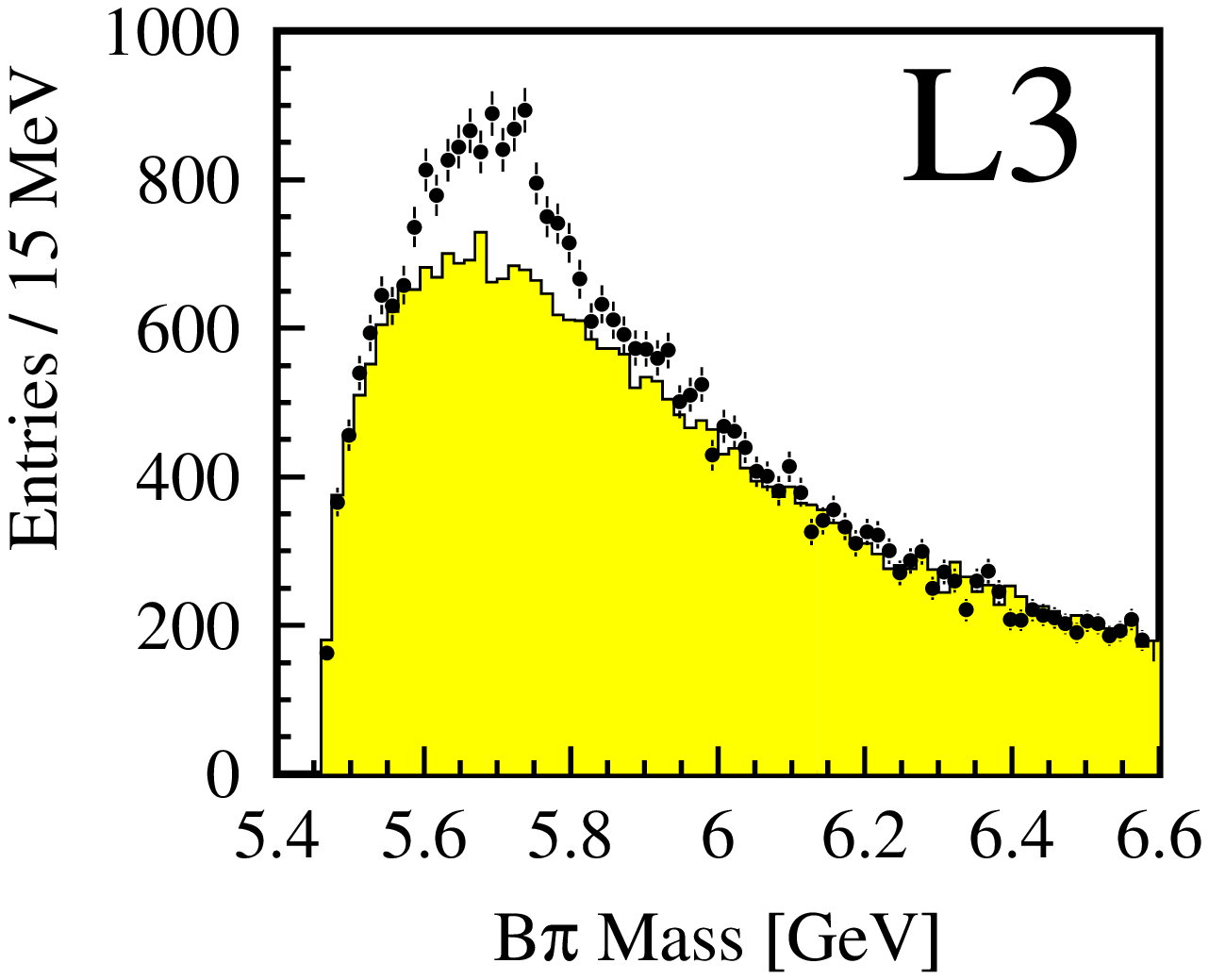,width=.49\linewidth}
    \epsfig{file=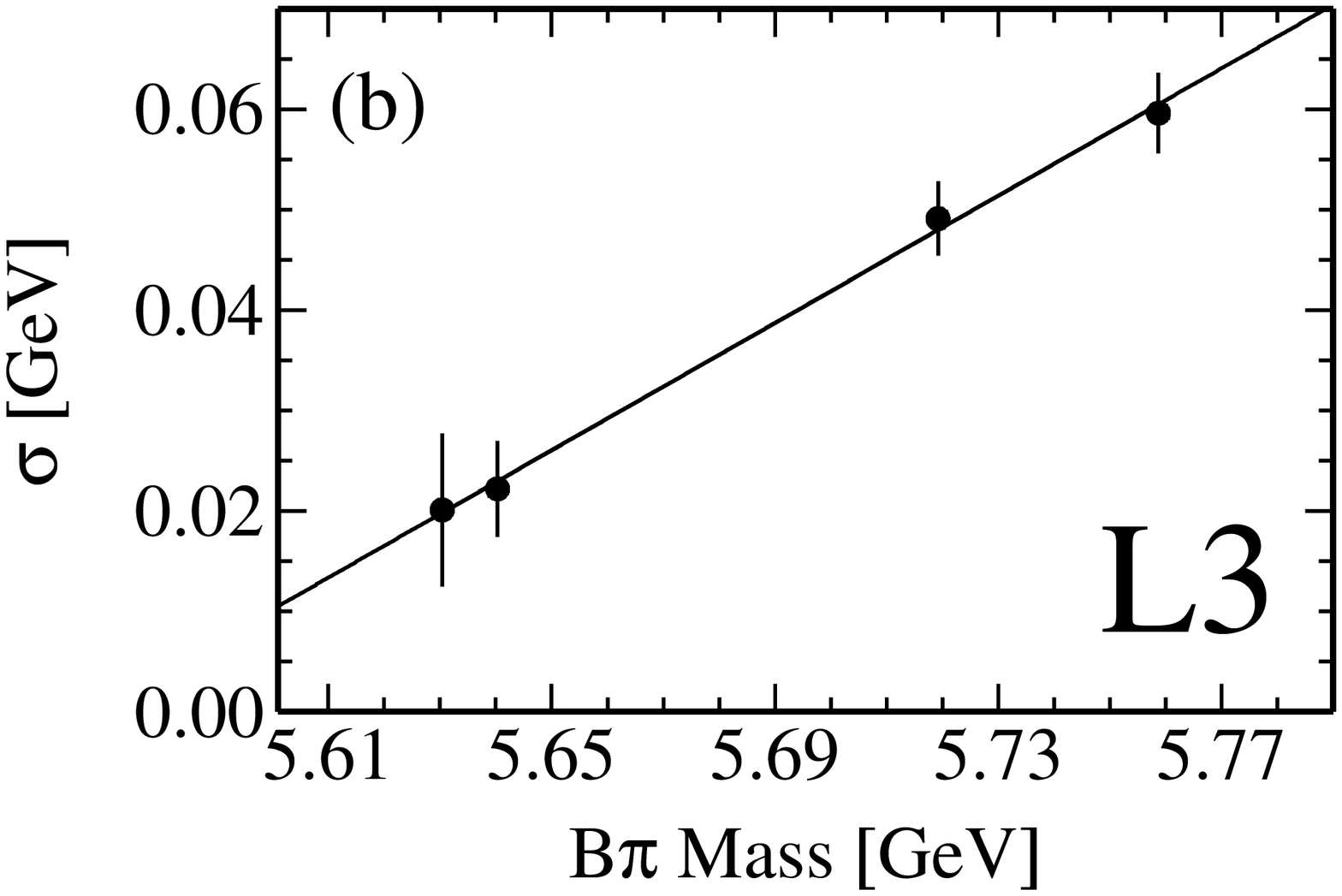,width=.49\linewidth}}
  \end{center}
  \caption[]{ (a) Mass spectrum for selected $\Bmeson\pi$ pairs.  The dots are
            data and the shaded histogram represents the expected background
            from Monte Carlo normalized to the sideband region $6.0-6.6~\GeV$.
	    (b) Linear fit to the extracted $\Bmeson\pi$ mass resolution 
	    for Monte Carlo signal components
            generated at four different $\Bdstar$ mass values.
           \label{fig:BpiMass}}
\end{figure}

Fig.~\ref{fig:BpiMass}a) shows the resulting $\Bmeson \pi$ invariant mass spectrum together
with the expected background from Monte Carlo. A clear signal  due $\Bdstar \to \BorBstar\pi$ 
decays is seen above the background which is well described by the simulation. Thus the background is parameterized by a
threshold function, the shape of which is determined from the Monte Carlo.

To resolve  the underlying structure of the signal, it is necessary
to unfold effects due to detector resolution.  
The dominant sources of uncertainty for the mass measurement
are, with about equal magnitude, the angular and energy resolution of the B meson.
The dependence of the $\Bmeson\pi$ mass resolution on the 
$\Bdstar$ mass is studied by simulating signal events at four different values
of $\Bdstar$ mass and Breit-Wigner width.  Each signal $\Bmeson\pi$ mass
distribution is then fit to 
a Breit-Wigner function convoluted with a Gaussian resolution.  The
Breit-Wigner width is fixed to its generated value and 
the Gaussian resolution is extracted from the fit and shown in Fig.~\ref{fig:BpiMass}b)
as a function of the  $\Bmeson\pi$ mass together with a linear parameterization.
The mass resolution is increasing with increasing  $\Bmeson\pi$ mass.

\begin{figure}[!htb]
  \begin{center}
      \mbox{\epsfig{file=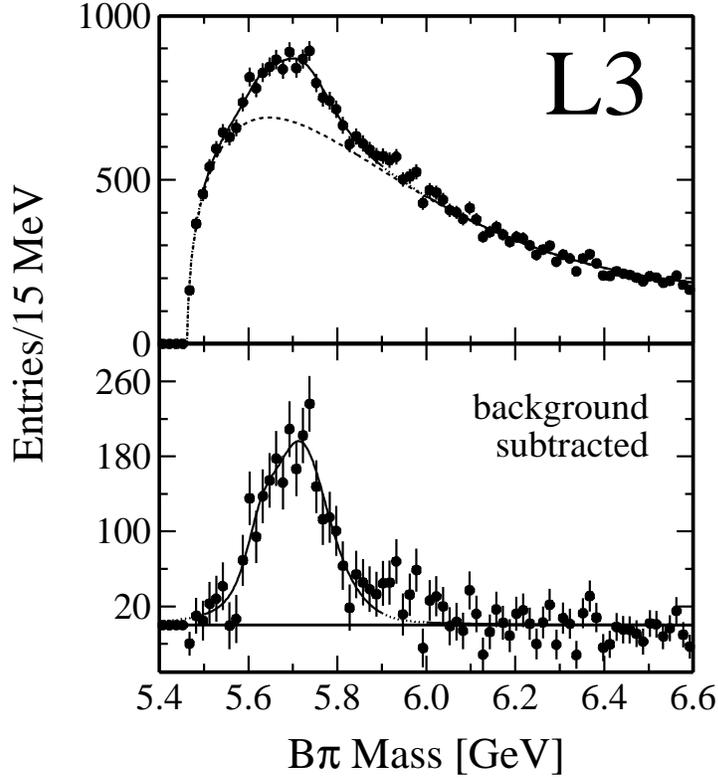,width=.7\linewidth}}
  \end{center}
  \caption{Fit to the data $\Bmeson\pi$ mass distribution with the five-peak
            signal function and the background function described in the text.
            \label{fig:Voigt}}
\end{figure}

The fit function for the signal consists of five Breit-Wigner mass peaks 
one for the each of the five decay modes allowed by spin-parity rules:
 $\Btwostar \to \Bmeson\pi, \Bstar \pi$, $\Bone \to \Bstar \pi$, 
$\Bonestar \to \Bstar \pi$, and $\Bzerostar \to \Bmeson \pi$.
Each  Breit-Wigner width is convoluted with the 
mass dependent Gaussian resolution. 
 No attempt is made to tag subsequent $\Bstar\rightarrow\Bmeson\gamma$ decays, as the
efficiency for selecting the soft photon is low.  
The relative production rate, and
the mass splittings and the relative width within each doublet are
constrained to the predictions from HQET (see Table~\ref{tab:decays}).

The $\Bmeson\pi$ invariant mass distribution fit with the signal and background function
described above is shown in Fig.~\ref{fig:Voigt}. 
The results of the fit
provide the first measurements of
the masses and decay widths of the $\Btwostar$ ($j_q=3/2$) and $\Bonestar$
($j_q=1/2$) mesons:
\begin{eqnarray*}
  M_{\Btwostar}      & = & (5770 \pm 6 \stat \pm 4 \syst)~\MeV \\
  \Gamma_{\Btwostar} & = & (21 \pm 24 \stat \pm 15 \syst)~\MeV \\
  M_{\Bonestar}      & = & (5675 \pm 12 \stat \pm 4 \syst)~\MeV \\
  \Gamma_{\Bonestar} & = & (75 \pm 28 \stat \pm 15 \syst)~\MeV \quad .
\end{eqnarray*}
A total of $2652 \pm 232$ events that occupy
the signal region correspond to a relative production rate $f_{\Bdstar}$ for 
all $L=1$ spin states of
\begin{eqnarray*}
   f_{\Bdstar}
    = 0.39 \pm 0.06 \stat \pm 0.06 \syst \quad .
\end{eqnarray*}
Systematic errors are mainly due to the modelling of the background, the limited knowledge of the signal function and the mass constraint within the doublets.

These results disfavor recent theoretical models proposing spin-orbit
inversion \cite{Isgur,Ebert}, but do agree well with several earlier
models \cite{Gronau,Eichten,Falk} and provide strong support for HQET.

\section*{\Bsubc\ Studies}

DELPHI, ALEPH, and OPAL have published searches for \Bsubc\ mesons in $Z$ decays\cite{BsubcADO}.
No signals have been found.
Table~\ref{tab:Bsubc} shows the number of candidate events and the obtained upper limits on the 
production rates ${\cal B}( Z \to \Bsubc X) \times 
{\cal B}(\Bsubc \to  J/\psi \pip,  J/\psi \ell^+ \nu, J/\psi \pip \pim \pip)$.
\begin{table}[!htb]
    \caption{ \Bsubc\ studies at LEP.\label{tab:Bsubc}}
  \begin{center}
\begin{tabular}{lccclll}
{\bf Decay mode} & \multicolumn{3}{c}{{\bf Candidates}} & \multicolumn{3}{c}{{\bf Prod. Limit $[10^{-5}]$ at 90\% CL}} \\
~& DELPHI & ALEPH & OPAL & DELPHI & ALEPH & OPAL \\ 
\tableline
{ $\Bsubc \to J/\psi \pip$}
& 1	& 0	& 2	& 10.5 to 8.4	& 3.6	& 10.6 	\\
{ $\Bsubc \to J/\psi \ell^+ \nu$} 
& 0	& 2	& 1	& 5.8 to 5.0	&5.2	& 6.96 	 \\
{ $\Bsubc \to J/\psi \pip \pim \pip$} 
& 1	& -	& 0	& 17.5		& -	& 5.53	\\ [1mm]
\end{tabular}
  \end{center}
\end{table}
\vspace*{-2mm} The  3 $\Bsubc \to J/\psi \pip$ candidates are consistent with a 
background estimate of 2.3 expected events. A fit to the mass yields  the following values:
$M_{J/\psi \pip} = 6.342 \pm 0.027$~\GeV\ (DELPHI) and $M_{J/\psi \pip} = 6.32 \pm 0.06$~\GeV\ (OPAL average), respectively.
Predictions for the \Bsubc\ mass are in the range 6.24 to 6.31 \GeV.
The CDF experiment at the Tevatron has recently reported the observation of the \Bsubc\ meson in the decay channel  $\Bsubc \to J/\psi \ell^+ \nu$\cite{BsubcCDF}. They find
 $20.4 \; ^{+6.2}_{-5.8}$ events and obtain a 
mass value of $M(\Bsubc) = 6.40 \pm 0.39 \pm 0.13$ \GeV.

\section*{\Ddstar\ Spectroscopy}

\begin{table}[!b]
    \caption{ \Ddstar\ production fractions ${\cal B}$ [\%]. \label{tab:Ddstar}}
\begin{center}
\begin{tabular}{lccc}
 {\bf Production mode} & {\bf OPAL} & {\bf DELPHI} & {\bf ALEPH} \\
\tableline
$b \to \Done^0$		& $5.0 \pm 1.4 \pm 0.6$	& $2.0 \pm 0.6$ & $2.3 \pm 0.7$ \\
$b \to \Dtwostarzero$	& $4.7 \pm 2.4 \pm 1.3$ & $4.8 \pm 2.0$	& $ < 2.0 $ \\
$c \to \Done^0$		& $2.1 \pm 0.7 \pm 0.3$ & $1.9 \pm 0.4$ & $1.6 \pm0.5$ \\
$c \to \Dtwostarzero$	& $5.2 \pm 2.2 \pm 1.3$ & $4.7 \pm 1.3$ & $ 4.7 \pm 1.0$ \\
$c \to \DdstarSone$	& $1.6 \pm 0.4 \pm 0.3$	& -	& $ 0.77 \pm 0.20 \pm 0.08$ \\
$c \to \DdstarStwo$	& -			& -	& $1.3 \pm 0.5 \pm 0.2 $ \\
$b \to \DdstarSone$	& -			& -	& $1.1 \pm 0.3 \pm 0.2 $ \\
$b \to \DdstarStwo$	& -			& -	& $2.2 \pm 0.8 \pm 0.5 $ \\ [1mm]
\end{tabular}
\end{center}
\end{table}

During the last year several new results on \Ddstar\ production have been presented by the 
LEP collaborations. \Ddstarzero\ mesons are fully reconstructed in  the decay chain
$\Ddstarzero \to \Dstarp \pim$, $\Dstarp \to \Dzero \pip$, $\Dzero \to \km \pip$. 
High momentum $\Ddstarzero$ together with short decay lengths  are selected  to obtain 
 $c \bar c$ enriched samples whereas B and D meson vertexing is used to select 
$b \bar b$ enriched samples. Table~\ref{tab:Ddstar} shows the results 
for the \Ddstar\ production fractions
measured by OPAL, DELPHI, and ALEPH\cite{DdstarOPAL,DdstarDELPHI,DdstarALEPH}.

\begin{figure}[thb]
  \begin{center}
  \mbox{\epsfig{figure=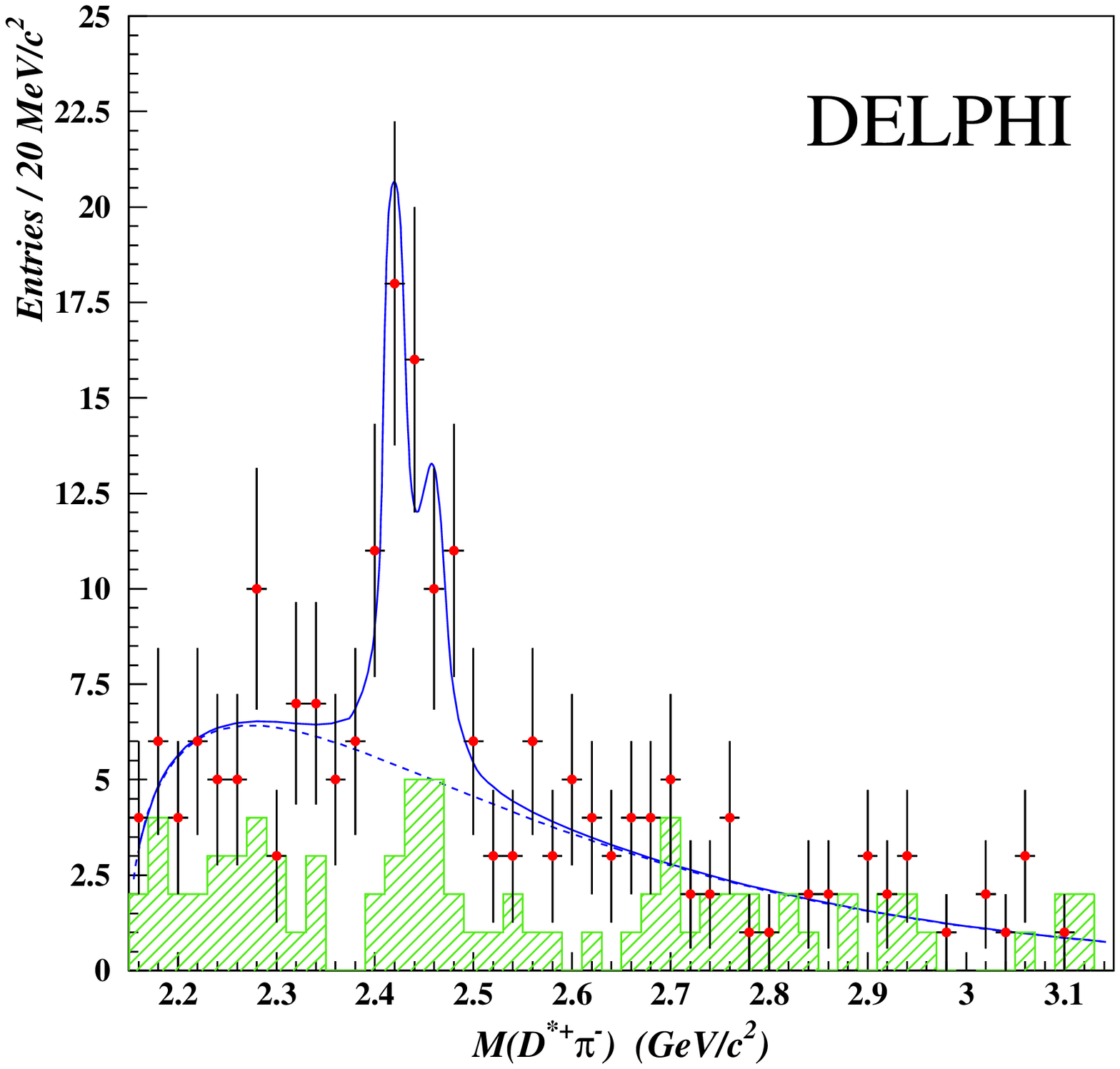,width=.48\linewidth}
  	\epsfig{figure=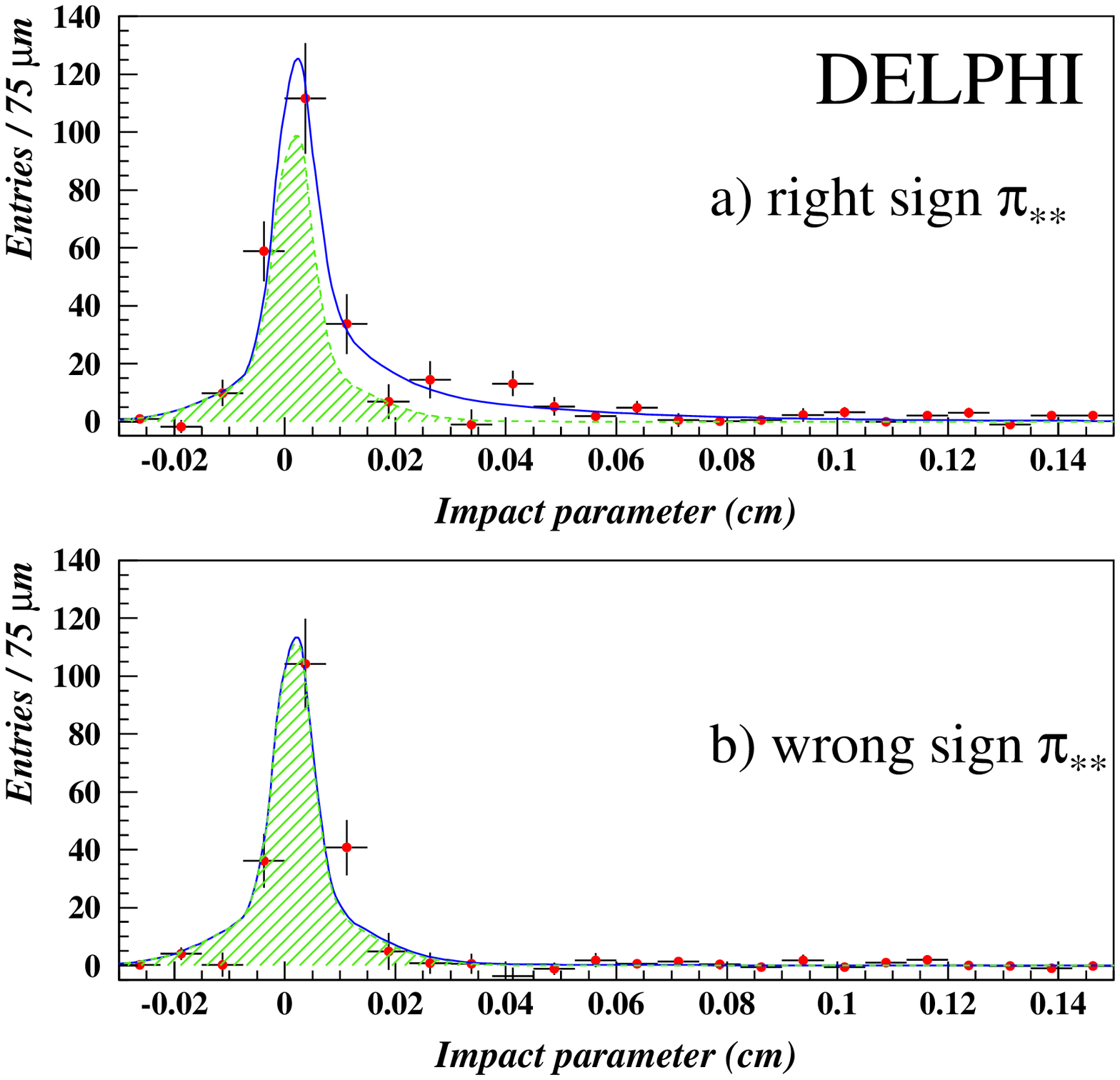,width=.48\linewidth}}
  \end{center}
\caption{(a) Experimental  $\Dstarp \pim$ invariant mass distribution for 
	$\Dstarp \pim \ell^-$ events. 
	The solid line is a fit to the narrow \Ddstar\ states plus background.
	The wrong sign   $\Dstarp \pim \ell^+$ candidates are shown in the hatched histogram. 
	(b) Impact parameter relative to the primary event vertex for right charge and 
	wrong charge  $\pi_{**}$ candidates. The fit is described in the text.
  \label{fig:Ddstarlnu}}
\end{figure}

DELPHI has also presented a new $\Bmeson \to \Ddstar \ell \nu$ analysis\cite{Ddstarlnu}.
In semileptonic events, the decay chain $\Dstarp \to \Dzero \pip$,
$\Dzero \to \km \pip, \km \pip \pim \pip, \km \pip (\pio)$ is fully reconstructed.
The $\Dstarp$ candidates are then combined with opposite sign $\pim$
and the $\Dstarp \pim$ mass, shown in Fig.~\ref{fig:Ddstarlnu}a),
is fit to the narrow \Ddstar\ states, resulting in the following branching fraction:
${\cal B}(B^- \to \Done^0 \ell^- \bar \nu) = 0.72 \pm 0.22  \pm 0.13 \%.$

A fit to the impact parameter distribution of the bachelor pion $\pi_{**}$ stemming from the 
$\Ddstar \to \Dstar \pi$ transition for right (unlike)-sign and wrong (like)-sign combinations,
as shown in  Fig.~\ref{fig:Ddstarlnu}b),
allows to extract the following branching fraction  
${\cal B}(B^- \to \Dstarp \pim \ell^- \bar \nu) = 1.15 \pm 0.17  \pm 0.14 \%$ 
where the signal comprises narrow and wide \Ddstar\ resonances plus
non-resonant $\Dstarp \pim$ combinations.
These results are in agreement with previous LEP and CLEO measurements.

\section*{\Dstarpr\  Studies}

\begin{figure}[thb]
  \begin{center}
\mbox{\epsfig{figure=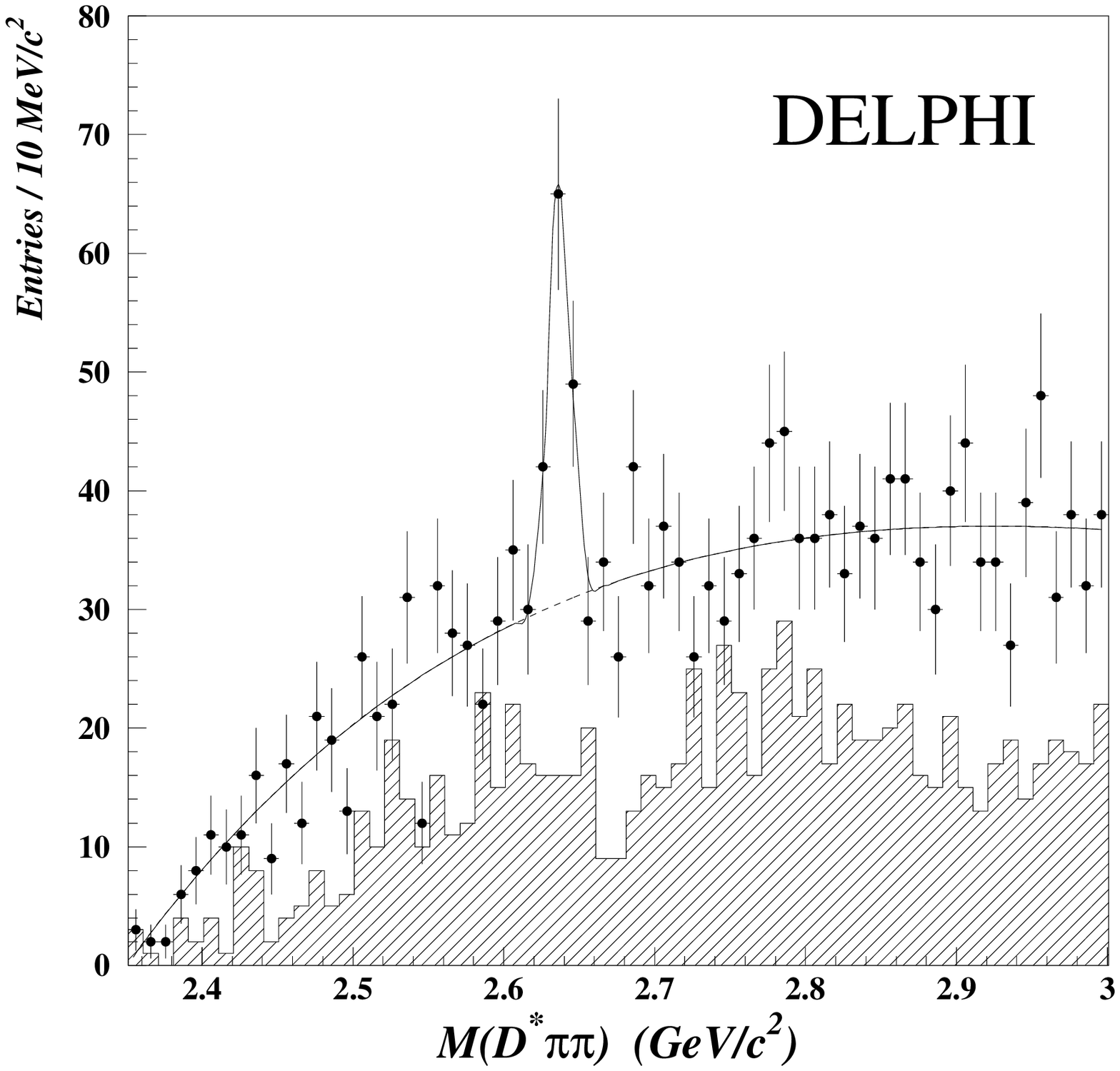,width=.55\linewidth}\epsfig{figure=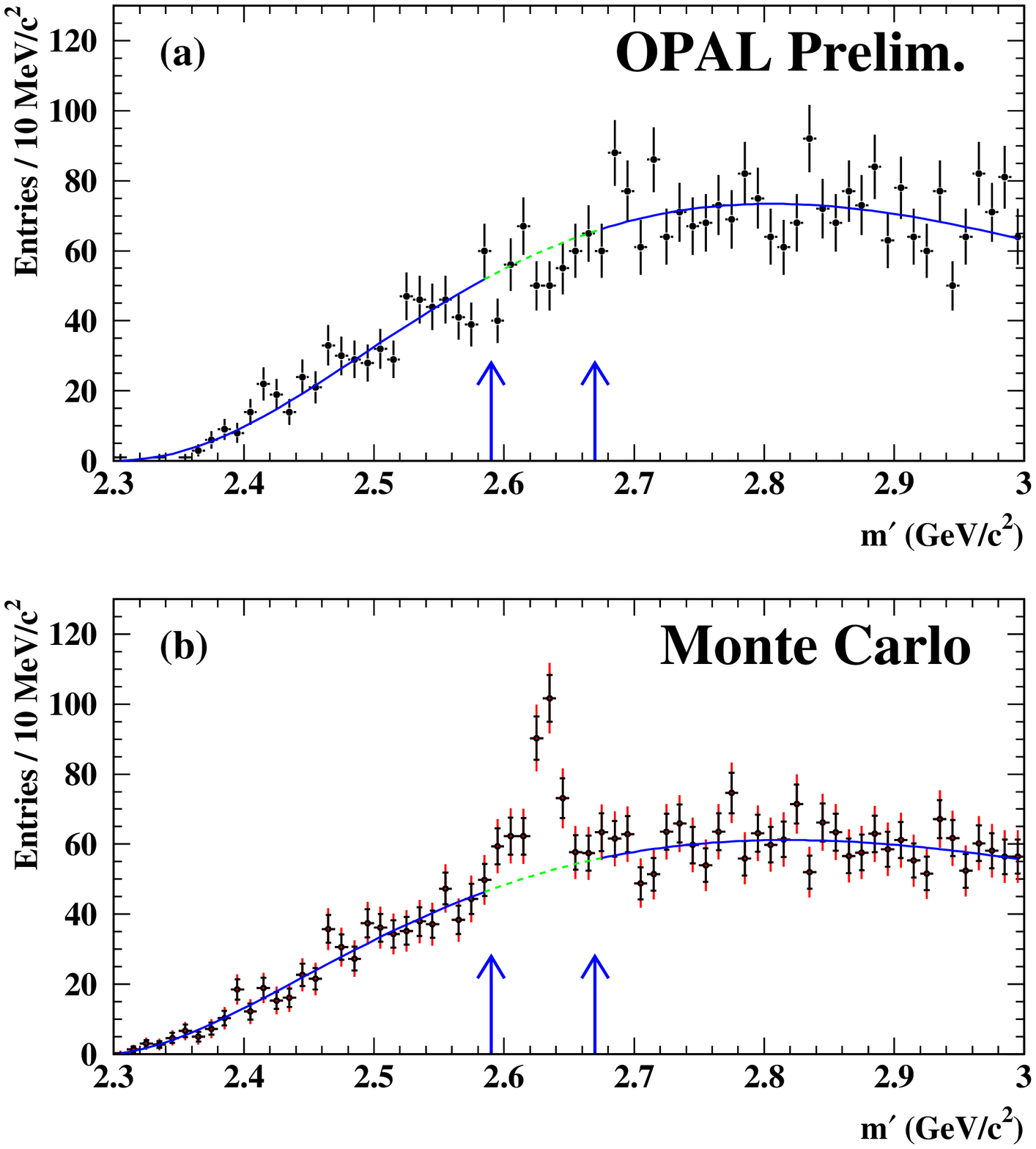,width=.45\linewidth}}
  \end{center}
\caption{(a) DELPHI invariant  mass distributions 
$\Dstarp \pip\pim$ (dots) and  $\Dstarp \pim\pim$ (hatched histogram). (b) OPAL $\Dstarp \pip\pim$
mass distribution for data and Monte Carlo. A  DELPHI-like signal has been added in the simulation.
  \label{fig:Dstarpr}}
\end{figure}

DELPHI has recently reported an excess of events  in the  $\Dstarp \pim \pip$ mass spectrum
as shown in Fig.~\ref{fig:Dstarpr}a)\cite{DstarprDELPHI}. 
The fit yields $N = 66 \pm 14$ events, corresponding to a production rate 
$f_{\Dstarpr}/f_{\Ddstar} = 0.49 \pm 0.18 \pm 0.10$, a mass $M = 2637 \pm 2 \pm 6$ \MeV\ 
and a width consistent with the experimental resolution.
This mass value is consistent with predictions of a radial excited \Dstarpr\ meson.

OPAL has performed a similar analysis\cite{DstarprOPAL}. The resulting   $\Dstarp \pim \pip$ mass spectrum is shown
in Fig.~\ref{fig:Dstarpr}b) for data and Monte Carlo events, where a DELPHI-like signal has been added in the simulation. No excess is seen in the data ($N < 32.8$ at 95 \% CL)
corresponding to a limit on the production rate of 
 $f_{\Dstarpr}/f_{\Ddstar} < 0.21$ at 95 \% CL,
thus not confirming the DELPHI result.
CLEO also has examined their  $\Dstarp \pim \pip$ mass spectrum and does not confirm 
the DELPHI evidence\cite{DstarprCLEO}.

\section*{Acknowledgements}
I wish to thank my colleagues from the other LEP collaborations for  
providing me with the results and the figure files. 
I also thank S. Goldfarb for his help when preparing this presentation.


\end{document}